\documentclass[preprint,amsmath,amssymb,showpacs]{revtex4} 

\usepackage{graphicx}
\usepackage{dcolumn}
\usepackage{bm}
\usepackage[cp1250]{inputenc}
\usepackage{ifthen} 
\usepackage{wrapfig}

\begin{document}

\title{Magnetoelectric correlations in BiMnO$_3$ whithin Landau theory: comparison with experiment}

\author{O.~Howczak}
\email[Electronic mail: ]{olga.howczak@uj.edu.pl}
\altaffiliation{Marian Smoluchowski Institute of Physics, Jagiellonian University, Reymonta 4, 30-059 Krak\'ow, Poland}

\author{J.~Spa\l ek}
\email[Electronic mail: ]{ufspalek@if.uj.edu.pl}
\altaffiliation{Marian Smoluchowski Institute of Physics, Jagiellonian University, Reymonta 4, 30-059 Krak\'ow, Poland}
\date{\today}



\begin{abstract}
	We discuss a simple phenomenological Landau theory of phase transitions
	with two coupled single-component order parameters and compare the results with available experimental data. The model  corresponds to the case of a ferroic system, in which ferromagnetic
	and ferroelectric transitions originally occur at temperatures $T_M$ and $T_f$, respectively.
	For $T_f>T_M$ the magnetoelectric
	coupling strongly renormalizes  the magnetic transition temperature, $T_M\rightarrow T_{RM}$ (with $T_{RM}>>T_M$), as well as generates an additional anomaly  in ferroelectric subsystem $T_{RM}$. Full susceptibility tensor has also been determined. The concept of \textit{Arrot plot} is replaced by the  \textit{Arrot planes} which appear when both types of order coexist. The results 	are in good overall agreement with experimental data for the ferroelectromagnetic BiMnO$_3$. We also estimate the contribution of Gaussian 	fluctuations of both order parameters, that lead to corrections to the mean-field specific heat. 
	Those corrections are still insufficient even though other quantities agree quite well with experiment.  
We calculate the temperature dependence of the coherence length for both types of order as well.   
\end{abstract}

\maketitle

\section{Introduction}

Multiferroics  are  materials, in which at least two types of order coexist. For example, (anti)ferroelectricity and
(anti)ferromagnetism can take place simultaneously \citep{eerenstein,khomskii}. 
Multiferroicity of frustrated magnets, in which magnetism
and ferroelectricity coexist with gigantic magnetoelectric
coupling, has attracted an interest  due to challenges to many-body theory, as well as by discoveries of new phenomena with a promice for potential applications as transducers, actuators, and sensors \citep{srinivasan,hur, zheng}.
Quite a few of these multiferroics are manganites, in which
the magnitude of the spin of the Mn$^{+3}$ ion is large and hence
may be treated semiclassically.

The perovskites like  AMnO$_3$ has been widely studied in this context due to the coexistence of ferroelectric and magnetic order  in some of them \citep{lee, daj, kenzelmann}. In BiMnO$_3$ the nature of the A$^{3+}$ ion is central
to determine the structural, ferroelectric, and magnetic properties of this system \citep{lancaster}.
With the help of first principle calculations, Hill et al. \citep{hill} provided the reasons why we observe so few ferroelectric magnets and predicted the existence of ferroelectricity in BiMnO$_3$. In ferroelectrics such as SrTiO$_3$ it is usually
driven by a hybridization of empty $3d^0$ transition metal orbitals with occupied
$2p$ orbitals of the octahedrally coordinated oxygen ions.
The appearance of magnetic moment in turn, requires partial occupancy of the $3d$ orbitals. So, in materials such as BiMnO$_3$, the coexistence of Mn$^{3+}$ ions ($3d^4$ configuration) with $6s^2$ lone electron pairs due to the Bi$^{3+}$ ions can lead to the 
coexistence of magnetic order with electric polarization at
low temperature \citep{khomskiiP}. 

BiMnO$_3$ has been the subject of considerable interest, mainly due to its 
structural simplicity. From magnetic point of view, Mn$^{3+}$ ion in this case has magnetic moment 
of $3.6\mu_{B}$, a value close to ground-state $4\mu_{B}$, induced by the Hund's rule coupling. In the octahedral environment
the electronic configuration is $t^3_{2g}e^{1}_g$. From electrical point of view BiMnO$_3$ is an insulator \citep{kaul}. 

A detailed structural study of  
BiMnO$_3$ suggests \citep{belik, montanari} that the material has a highly
distorted perovskite structure (centrosymmetric space group C2/c) which is incompatible with the existence of ferroelectricity.
In spite of this, there is an experimental evidence of ferroelectric order in a BiMnO$_{3}$ polycristaline samples \citep{santos, kimura, belik}. 
The discussion concerning the origin of the ferroelectricity in this compound is still controversial. 

We present  a simple phenomenological approach useful 
in describing systems like BiMnO$_3$ with two coupled order parameters, as well as compare the results with experiment.
This is to show to what extent a simple Landau-type approach can account for the experimental
results in a quantitative manner. Our task is related to an even more basic question to what  extent ferroelectric order (appearing first at much higher temperature
$T_f$) suppresses the magnetic fluctuations near the corresponding transition temperature $T_M<<T_f$. In such  situation the mean-field description of the magnetic phase should be at least semiquantitatively correct, as we demonstrate below. 

The structure of the paper is as follows.
In Sections II and III we formulate the Landau-theory with two single-component, spatially homogeneous order parameters
$P$ and $M$ (representing the ferroelectric and the ferromagnetic types of order, respectively) coupled 
via a phenomenological term of the type $-|\gamma| (PM)^2$. This coupling leads to a renormalized 
magnetic transition temperature. In Section III we also compare the results obtained  with the data available 
 for BiMnO$_3$, as well as introduce a new concept of \textit{Arrot plates}, which can be used in systems with two coupled order parameters
 in a ordered state.  In Section IV  we include Gaussian fluctuations for the coupled 
system, and subsequently compare the results with experimental data concerning the temperature dependence 
of the specific heat near the low-temperature magnetic transition. We summarize our results and provide an outlook in Section V.

\section{Mean Field Approximation: Landau Approach}

The way to describe the coupling between magnetism and dielectricity in multiferroics  was proposed by
 Smolenskii \citep{smol}, who explained the origin of the anomaly in the dielectric constant in a  ferroelectromagnet within the framework of Landau theory of second-order phase transitions.
In the simplest case, the Landau free energy for a system with two coupled order parameters, $P$ and $M$ can be written as: 
\begin{eqnarray}
F(P,M,T)&=&F_0(T)+\frac{a_0(T-T_M)}{2} M^2+ \frac{b}{4}M^4 \\\nonumber
&+& {}  \frac{\alpha_0(T-T_f)}{2} P^2+ \frac{\beta}{4}P^4 \\\nonumber
&+& {} \frac{\gamma}{2}(PM)^2- PE_a-MH_a,
\label{l}
\end{eqnarray}

where $E_a$ and $H_a$ are the applied electric and magnetic fields, respectively. Recently it was shown that  this kind of model can  discribe the phase transitions on a scale-free network as well \citep{pal} . 
As one can see, the system is characterized by two bare transition temperatures, $T_f$ and 
$T_M$ representing ferroelectric and ferromagnetic transitions, respectively. 
The form of the magnetoelectic coupling term $\propto \gamma (PM)^2$ can be explained by using general symmetry arguments. Namely, the onset of ferroelectric order requires the breaking of spatial inversion symmetry, whereas the appearance of a spontaneous magnetization  is connected with the breakdown of time reversal symmetry. The coupling term in (\ref{l}) obeys  those two conditions and allows for  ferroic order with simultaneous nonzero $P$ and $M$.

In further calculations we use a dimensionless form of the free energy expansion (\ref{DDF}) which is obtained from (\ref{l}) by dividing both sides of (\ref{l}) by the constant value $\frac{a_0^2 T_M^2}{b}$.
In effect, we obtain:  
\begin{eqnarray}
 \Delta F&=& \frac{1}{2}\left(\frac{T}{T_M}-1\right)\tilde M^2+\frac{1}{4}\tilde M^4\\\nonumber
  &&{} +\frac{1}{2}l\left(\frac{T_f}{T_M}\right)^2\left(\frac{T}{T_f}-1\right)\tilde P^2 + \frac{1}{4}l\left(\frac{T_f}{T_M}\right)^2\tilde P^4\\\nonumber 
&& {} +\frac{1}{2}\gamma_m\frac{T_f}{T_M}\tilde P^2\tilde M^2-\tilde{P}e-\tilde{M}h,
\label{DDF}
\end{eqnarray}
where:
$\Delta F=\frac{b}{a_0^2T_M^2}\left[F( P,  M, T)- F_0(T)\right]$, 
$h=\frac{b^{\frac{1}{2}}}{(a_0 T_M)^\frac{3}{2}} H_a\equiv s_h H_a$,
$l=\frac{\alpha_0^2 b}{a_0^2\beta}$, $e=\frac{b}{a_0^2 T_M^2}\sqrt{\frac{\alpha_0 T_f}{\beta}} E_a$,  
$\tilde{M}=\frac{M}{m_0}$, $m_0=\sqrt{\frac{a_0 T_M}{b}}$,  
$\tilde{P}=\frac{P}{p_0}$, $p_0=\sqrt{\frac{\alpha_0 T_f}{\beta}}$, 
$\gamma_{m}=\frac{\gamma \alpha_0}{\beta a_0}$.  

Within this formulation $\frac{T_f}{T_M}$, $l$  , and $\gamma_m$ are the material parameters.
In this case, the bare magnetic transition  temperature $T_M$ is assumed as always
nonzero, whereas the ferroelectric temperature is $T_f\geq0$.

Taking derivatives  of  (\ref{DDF}) with respect to $\tilde{M}$ and $\tilde{P}$, respectively,  we obtain the following system of nonlinear equations:
\begin{widetext}
\begin{eqnarray}
(T/T_M-1)\tilde{M}+\tilde{M}^3+\gamma_m\frac{T_f}{T_M}\tilde{M}\tilde{P}^2-h& = & 0, \\
l\left(\frac{T_f}{T_M}\right)^2\left(T/T_f-1\right)\tilde{P}+l\left(\frac{T_f}{T_M}\right)^2\tilde{P}^3+\gamma_m \frac{T_f}{T_M}\tilde{M}^2\tilde{P}-e& = & 0.        
\label{u}
\end{eqnarray}
\end{widetext}
In the limit of zero electric field, we obtain the following relation 
between $\tilde{P}$ and $\tilde{M}$ from (\ref{u}):
\begin{equation}
\tilde{P}=\pm \sqrt{1- \frac{T}{T_f}- \frac{\gamma_m}{l}\frac{T_M}{T_f}\tilde{M}^2}.
\label{p}
\end{equation}  
Substituting this expression to (3) we have the equation for magnetization
in the usual form:
\begin{equation}
A(T) \tilde{M} +B \tilde{M}^3 -h=0,
\label{m}
\end{equation}
with
\begin{equation}
A\equiv A(T)= \frac{T}{T_M}(1-\gamma_m) +\gamma_m \frac{T_f}{T_M}-1,
\label{AA}
\end{equation}
and
\begin{equation}
B=1-\frac{\gamma_m^2}{l}.
\end{equation}

One sees that the effective magnetic transition temperature $T_{RM}$ 
is renormalized by the magnetoelectrical coupling. Explicitly, since the renormalized 
transition temperature is determined  from the condition $A(T_{RM})=0$, this 
yields:

\begin{equation}
T_{RM}=\frac{1-\gamma_{m}\frac{T_f}{T_M}}{1-\gamma_{m}}\cdot T_M.
\end{equation}
\label{Trc}

This is one of the interesting results. Namely, the renormalization is strong because of negative value of coupling constant $\gamma_m$.  Furthermore, the
renormalization of $T_f$ does not appear if $T_f>T_M$.
In the case of $T_f=T_M$ the coupling would not change 
the critical temperatures and no renormalization would occur.

The explicit stable solutions of equation (\ref{m}) for $\tilde{M}$ in the case $e=0$ are:
\begin{widetext}
\begin{equation} 
\tilde{M}=\left\{ \begin{array}{cccc}
-\frac{\left(\frac{2}{3}\right)^\frac{1}{3}A}{\left(9B^2h+\sqrt{3}\sqrt{4A^3B^3+27B^4h^2}\right)^\frac{1}{3}}+\frac{\left(9 B^2h+\sqrt{3}\sqrt{4A^3B^3+27B^4h^2}\right)^\frac{1}{3}}{2^\frac{1}{3} 3^\frac{2}{3} B}, &&  for\ h>0  \\
\frac{\left(\frac{2}{3}\right)^\frac{1}{3}A}{\left(\sqrt{3}\sqrt{4A^3B^3+27B^4h^2}-9B^2h\right)^\frac{1}{3}}-\frac{\left(\sqrt{3}\sqrt{4A^3B^3+27B^4h^2}-9B^2h\right)^\frac{1}{3}}{2^\frac{1}{3} 3^\frac{2}{3} B},  && for\ h<0
\end{array}
\right\}
\label{80}
\end{equation} 
\end{widetext} 

We apply the solution obtained  above to the discussion of selected magnetic and dielectric properties of BiMnO$_3$.

\section{Application to $BiMnO_3$}
\subsection{ Magnetic properties}
\begin{figure}[ht]
\centering
\includegraphics[width=90mm]{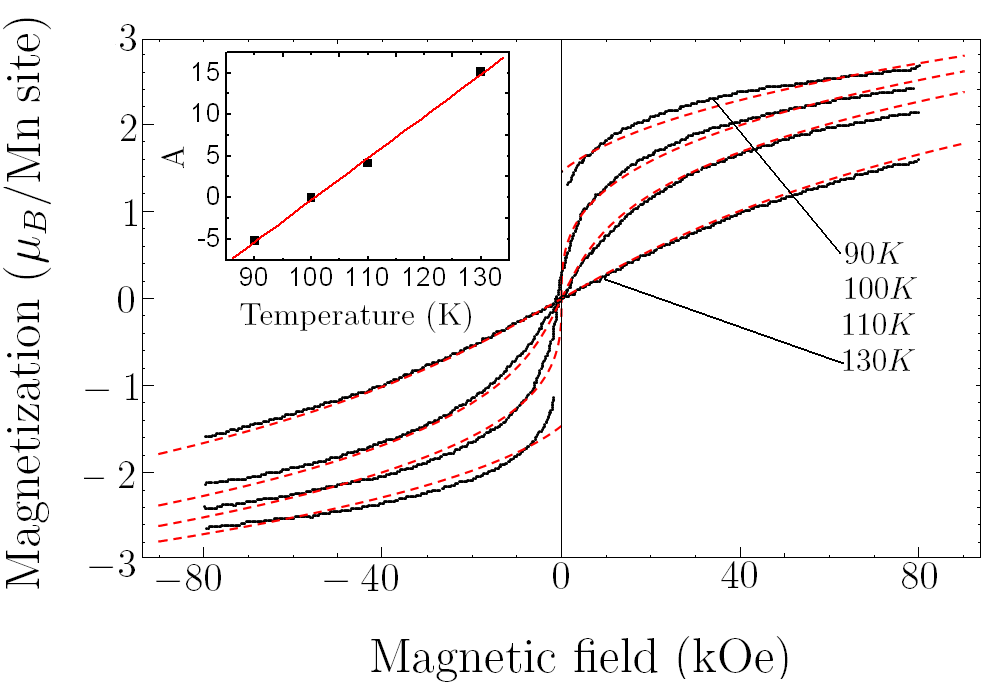}
\caption{\label{fig:mhh}(Color online). Isothermal magnetization as function of a magnetic field at various temperatures for BiMnO$_3$ \citep{kimura} (solid line) and that from Eq. (\ref{m}) (dashed lines). Inset: $A(T)$ values for the temperatures marked. The fitting parameters are listed in Table.\ref{table:nonlin}.}
\end{figure}

To visualize the influence of the magnetoelectric coupling on the magnetic properties of BiMnO$_3$,
we fitted the temperature and applied magnetic field dependences of the magnetization, based on the data of Kimura \citep{kimura} and Chiba \citep{chiba}.
In Fig. \ref{fig:mhh} we display the fitted $M(H_a; T)$ curves near the critical 
temperature ($T_{RM}\approx100 K$). In the inset we plot the values of $A(T)$ obtained from the fitting: it is 
indeed a linear function of $T$, as obtained in (\ref{AA}).
By taking the value $T_f=760 K$ \citep{kimura, chiba} we have obtained the renormalized value
of $T_{RM}\approx100.5 K$, the bare Curie temperature $T_M=2.28 K$, the magnetoelectric coupling
constant $\gamma_m\approx-0.15 $, and $l\approx0.024$. One sees that the renormalization $T_{RM}/T_M$ is very large with the increase caused by the negative sign of the coupling  constant $\gamma$. We consider the coupling in case of BiMnO$_3$ to be large because  $|\gamma_m|\approx\sqrt{l}$ and as it can be seen from Eqs. (19-\ref{MP3}) it directly effects the rapid increase of electric polarization and magnetization in ordered state and is the greatest possible value for the coupling.

\begin{figure}[ht]
	\centering
		\includegraphics[width=90mm]{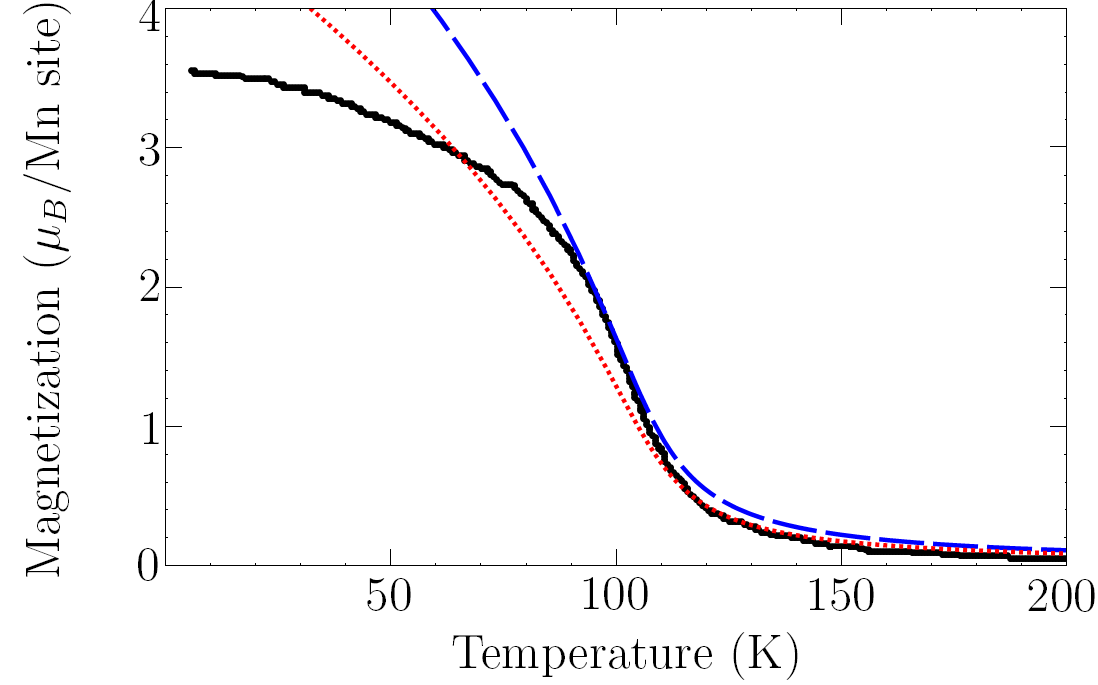}
\caption{(Color online).Temperature variation of magnetization of BiMnO$_3$ measured at 1T \citep{chiba} (solid line) and the fitted solution of (\ref{m}). Dotted line: The averaged (cf. Table.\ref{table:nonlin}) fitting parameters with $A(T)=-50.593 + 0.503	 T$, $B = 0.062$, $m_0 = 0.161\ \mu_B$/Mn site,  and $s_h=26.5\  T^{-1}$. A better fit (dashed line) can be obtained by a slightly different set of parameters: $A(T)=-49.926 + 0.507 T$, $B = 0.059$, $m_0 = 0.208\ \mu_B$/Mn site  and $s_h=24.6\  T^{-1}$.}	
\label{fig:mtn}
\end{figure}

In Fig. \ref{fig:mtn} we fitted the temperature dependence of magnetization \citep{chiba} in two ways:
first (dotted line), by taking the averaged values from Table.\ref{table:nonlin} and second, by a direct fitting i.e. changing slightly the averaged values (dashed line). Those slight changes are justified, as they are within statistical error. One should mention that data used in Fig. \ref{fig:mhh} was taken for a different sample to that of Fig. \ref{fig:mtn}.

From these two figures one sees, that the overall behavior of the magnetization near $T_{RM}$ is well reproduced 
by the mean field approach, particularly for $T\rightarrow T_{RM}$ and above. Hence, we parametrize  the dielectric and magnetoelectric susceptibility components in the same manner next.

\begin{table}[ht]
\caption{Fitting parameters obtained from the results of Fig. 1.}
\centering 
\begin{tabular}{| c | c | c | c | c |} 
\hline\hline 
T [K]& $A$ & $B$ & $m_0\  [\mu_B$/Mn site ] & $s_h\  [1/T]$ \\ [0.5ex]
\hline 
130 & 15.079(8) & 0.0623(4) & 0.1601(8) & 26.509(2) \\ 
110 & 4.12(63)    & 0.06(233) & 0.16(019) & 26.5(092) \\
100 & 0.00061   & 0.06(047) & 0.16(513) & 26.5(346) \\
90  & -5.1(848)   & 0.06(234) & 0.16(018) & 26.5(093) \\ [1ex] 
\hline \hline
\end{tabular}
\label{table:nonlin} 
\end{table}

\subsection{Susceptibility tensor}

Kimura et al. reported \citep{kimura} that  with increasing temperature the magnitude of the isothermal magnetocapacitance increases and exibits a maximum around $T_{RM}$. Whereas upon further increase (above $T_{RM}$) the magnetocapacitance  subsequently decreases. The authors claim that this phenomenon arises from the magnetization rotation in magnetic domains.
Such behavior can also be obtained from the simple Landau approach introduced here 
without involving any domain formation.
On application of external fields the system response in the ferromagnetoelectric state ($T<T_{RM}$) is described by the tensor:
\begin{equation}
 \hat{\chi}  =  \left( \begin{array}{cc}
\tilde{\chi}_e & \tilde{\chi}_{em}  \\
\tilde{\chi}_{me} & \tilde{\chi}_m  
\end{array} \right),
\end{equation} 

with:
\begin{equation}
\frac{\partial \tilde{M}}{\partial h}=\tilde{\chi}_m,\ \ \frac{\partial \tilde{P}}{\partial e}=\tilde{\chi}_e,\ \ 
\frac{\partial \tilde{M}}{\partial e}=\tilde{\chi}_{me},\ \ \frac{\partial \tilde{P}}{\partial h}=\tilde{\chi}_{em}.
\end{equation}
By assuming  $e\neq0$, we obtain the following equations for $\tilde{M}$ and $\tilde{P}$:
\begin{widetext}
\begin{eqnarray}
\frac{\partial(\Delta F)}{\partial \tilde{M}}&=& (T/T_M-1)\tilde{M}+\tilde{M}^3+\gamma_m \frac{T_f}{T_M}\tilde{M}\tilde{P}^2 -h=0,\\
\frac{\partial(\Delta F)}{\partial \tilde{P}}&=& l \left(\frac{T_f}{T_M}\right)^2\left(\frac{T}{T_f}-1\right)\tilde{P}+l\left(\frac{T_f}{T_M}\right)^2\tilde{P}^3+\gamma_m \frac{T_f}{T_M}\tilde{M}^2\tilde{P} -e=0.
\label{eq11}
\end{eqnarray}
\end{widetext}

After differentiating  (13) and (\ref{eq11}) with respect to both $e$ and $h$, we obtain a system of linear equations for the 
susceptibility components in the form: 
\begin{equation} 
 \begin{array}{cccc}
\tilde{B} \tilde{\chi}_m+\tilde{C} \tilde{\chi}_{em}&  & 1, \\
\tilde{B}\tilde{\chi}_{me}+\tilde{C} \tilde{\chi}_{e}& = & 0, \\
\tilde{A}\tilde{\chi}_e+ \tilde{C}\tilde{\chi}_{me}& = & 1 ,\\
\tilde{A}\tilde{\chi}_{em}+ \tilde{C}\tilde{\chi}_{m}& = & 0,
\end{array}  
\label{chi}
\end{equation} 

with:
$\tilde{A} \equiv \left(\frac{T_f}{T_M}\right)^2\left(\frac{T}{T_f}-1\right)+3l\left(\frac{T_f}{T_M}\right)^2\tilde{P}   ^2+\gamma_m\frac{T_f}{T_M}\tilde{M}^2$, $\tilde{B} \equiv \frac{T}{T_M}-1+3\tilde{M}^2+\gamma_m\frac{T_f}{T_M}\tilde{P}^2$, $\tilde{C} \equiv 2\gamma_m\frac{T_f}{T_M}\tilde{P}\tilde{M}$. 

Therefore the  solution of (\ref{chi})  takes the form:    
\begin{equation}
\tilde{\chi}_m =  \frac{\tilde{A}}{\tilde{A}\tilde{B}-\tilde{C}^2}, 
\label{c1}
\end{equation}

\begin{equation}
\tilde{\chi}_{e} =  \frac{\tilde{B}}{\tilde{A}\tilde{B}-\tilde{C}^2}, 
\label{ce}
\end{equation}
and
\begin{equation}
\tilde{\chi}_{me}=\tilde{\chi}_{em} =  \frac{\tilde{C}}{\tilde{C}^2-\tilde{A}\tilde{B}}. 
\label{c3}
\end{equation}
In the zero field case and for temperature $T < T_{RM}$, the corresponding  expressions for magnetization and polarization  are:

\begin{eqnarray}
\tilde{M}&=& \pm \sqrt{\frac{l}{l-\gamma_m^2}}\cdot\sqrt{1-\frac{T}{T_M}-\gamma_m\frac{T_f}{T_M}\left(1-\frac{T}{T_f}\right)},\\
\tilde{P}&=& \pm \sqrt{\frac{l}{l-\gamma_m^2}}\cdot\sqrt{1-\frac{T}{T_f}-\frac{\gamma_m}{l} \frac{T_M}{T_f}\left(1-\frac{T}{T_M}\right)}. 
\label{MP3}
\end{eqnarray}

For $T_{RM}<T<T_f$, i.e. in the ferroelectric state, we obviously have:
\begin{eqnarray}
\tilde{P}&=& \pm \sqrt{1-\frac{T}{T_f}},\\
\tilde{M}&=&0. 
\label{MP2}
\end{eqnarray}

\begin{figure}[ht]
	\centering
		\includegraphics[width=90mm]{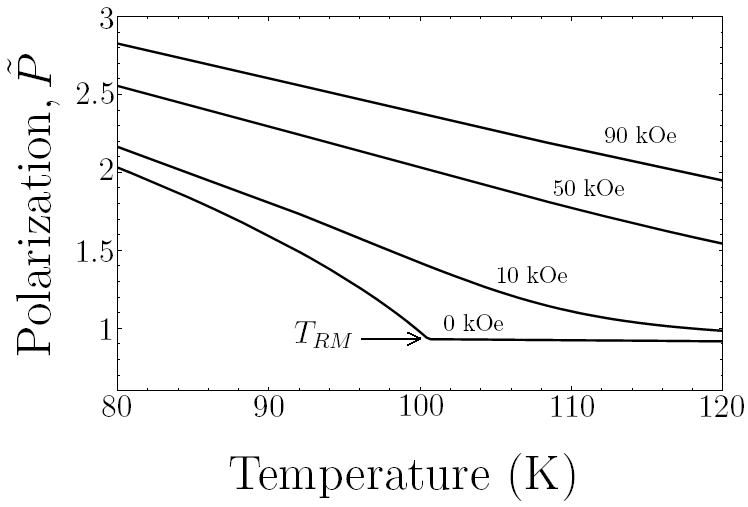}
		\caption{Polarization as a function of temperature for selected values of applied magnetic field. The parameters are the same as those used previously.} 	
	\label{fig:p}
\end{figure}

In Fig. \ref{fig:p} we plot the temperature dependence of the polarization in the vicinity of $T_{RM}$, i.e. at temperatures $T< T_f$. One sees that even though the ferroelectric ordering temperature 
is well above $T_{RM}$, the weaker-scale magnetic interaction significantly enhances the polarization.
This enhancement is also present on application of a magnetic field. This is a second (in addition to the  renormalization of $T_M$) important cross-effect correlating 
magnetic and electric properties in these systems. One should also note  that the electric polarization increases with increasing magnetic field,
as one may expect from the negative sign of the magnetoelectric coupling. Unfortunately, no experimental results are available to us to confront our findings with those for BiMnO$_{3}$.   

We now turn to the analysis of  the  susceptibility-tensor components. 
Substituting the values obtained above for the magnetization and polarization into (\ref{ce}), we 
obtain an explicit expression for the zero field electric susceptibility, namely
\begin{equation} 
\tilde{\chi}_e(0) = \left\{ \begin{array}{cccc}
\frac{1}{2}\frac{{T_M}^2}{T_f\left(l(T_f-T)+\gamma_m(T-T_M)\right)} && for\ T<T_{RM}, 
\\
\frac{{T_M}^2}{2lT_f(T_f-T)}  && for\ T>T_{RM}.
\end{array}
\right\}  
\label{chi80}
\end{equation} 

After calculating  $\tilde{\chi}_e(h)$, using the general  solutions of (\ref{80}), we plot:
\begin{eqnarray}
\Delta\epsilon(H_a)/\epsilon(0)&\equiv&[\epsilon(H_a)-\epsilon(0)]/\epsilon(0)\\
				&=& \frac{4\pi (\chi_e(H_a)-\chi_e(0)) }{1+4\pi\chi_e(0)}\\
				&=& \frac{\Delta\tilde{\chi}_e(h)}{\frac{\beta a_0^2 T_M}{4\pi \alpha_0 b T_f}+\tilde{\chi}_e(0)},
\end{eqnarray} 

\begin{figure}[ht]
	\centering
		\includegraphics[width=90mm]{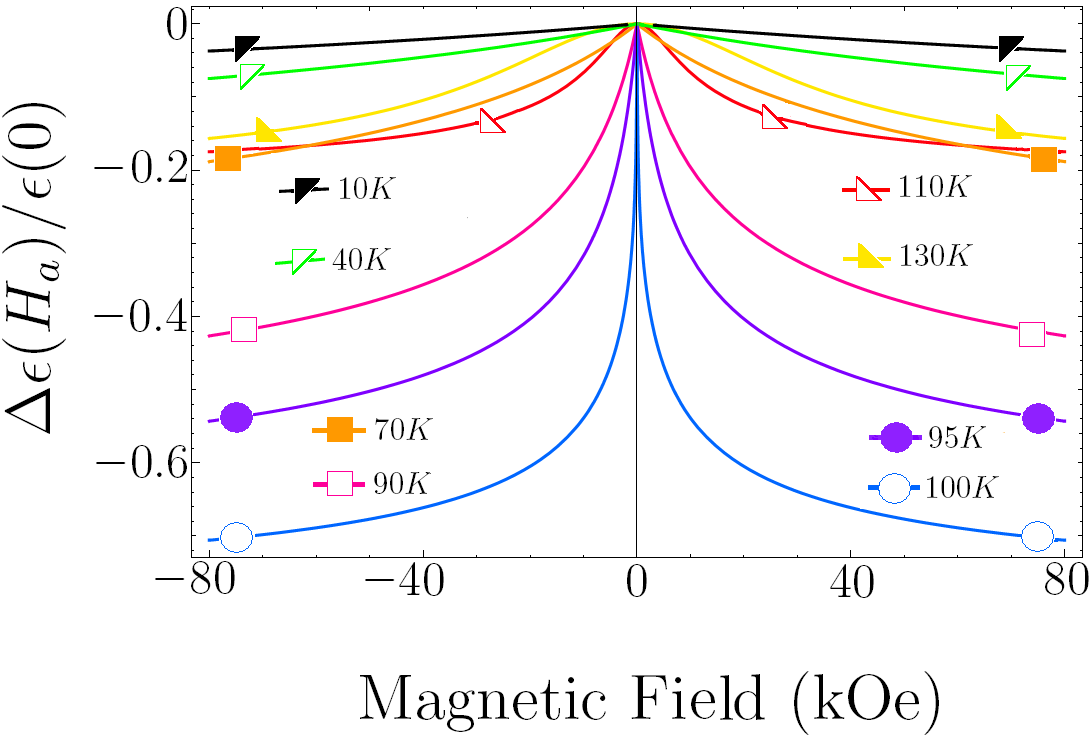}
	\caption{(Color online). Field-induced change in the dielectric constant as a function of an applied magnetic field for selected temperatures. We assume that $(\beta a_0^2 T_M)/(4\pi \alpha_0 b T_f)=0.0008$.}
	\label{fig:4}
	\end{figure}
shown in Fig. \ref{fig:4}. The curves obtained reflect  
the corresponding experimental data well \citep{kimura}. Though one should note, that the actual change of the dielectric constant is not as rapid as 
in our mean field approach.
 Also the rapid trend upward of the curves appears here above $100K$, whereas 
the respective changes of the data appears only above $110K$ \citep{kimura}. This difference 
is a clear sign of the nonzero value of the magnetization due to short-range 
correlation. This type of crossover behavior above $T_{RM}$ will appear 
in the specific heat data, as discussed in the next Section.  

\begin{figure}[ht]
	\centering
		\includegraphics[width=90mm]{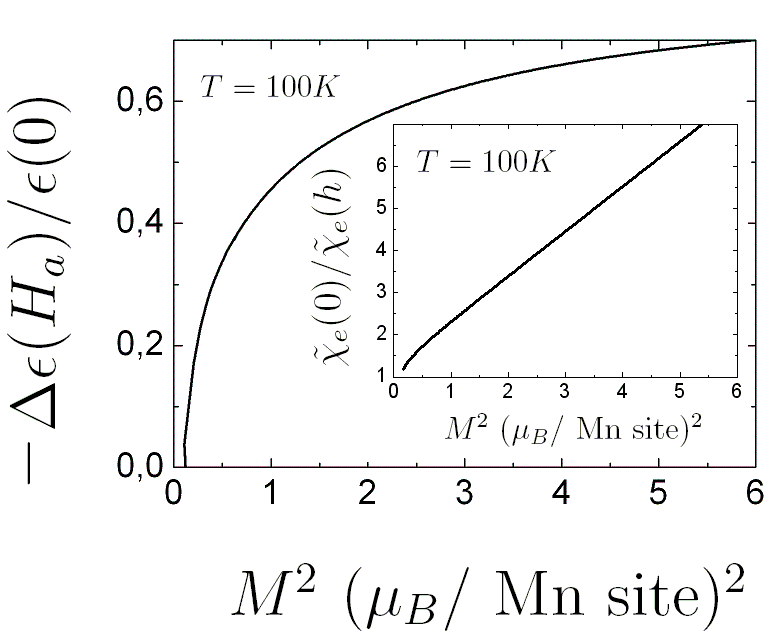}
		\caption{Field induced change in relative dielectric constant as a function of the square of the magnetization  at 100 K. We assume that $(\beta a_0^2 T_M)/(4\pi \alpha_0 b T_f)=0.0008$. Inset: Dependence of the inverse dielectric susceptibility vs. $M^2$. The data can be parametrized by the straight line: $\tilde{\chi}_e(0)/\tilde{\chi}_e(h) = 1,233 + 1,071 M^2 $.}
	\label{fig:dep}
\end{figure}

Experimentaly, the  susceptibility is a linear function of the squared magnetization for BiMnO$_{3}$  \citep{kimura}. In some papers \citep{kimura, zhong} this is 
rationalized on the basis  of Landau-theory yet this it is not the case.
In fact, in this phenomenological approach the inverse  susceptibility
is a linear function of the squared magnetization and it comes about from 
the renormalization of dielectric constant by the coupling. Explicitly this may be written \citep{chupits}  
$\Delta F\approx (\alpha_0(T-T_f)/2 + \gamma/2\ M^2)P^2 + ...=\chi_e^{-1}P^2+...$ .
In Fig. \ref{fig:dep} we plot the predicted Landau theory value of 
$\Delta\epsilon/\epsilon(0)$ as a function of the squared magnetization. The dependence is none linear.
In the inset of Fig. \ref{fig:dep} we show the inverse dielectric susceptibility as a function of the squared magnetization to be linear within the framework of Landau theory.
 
\begin{figure}[ht]
	\centering
		\includegraphics[width=90mm]{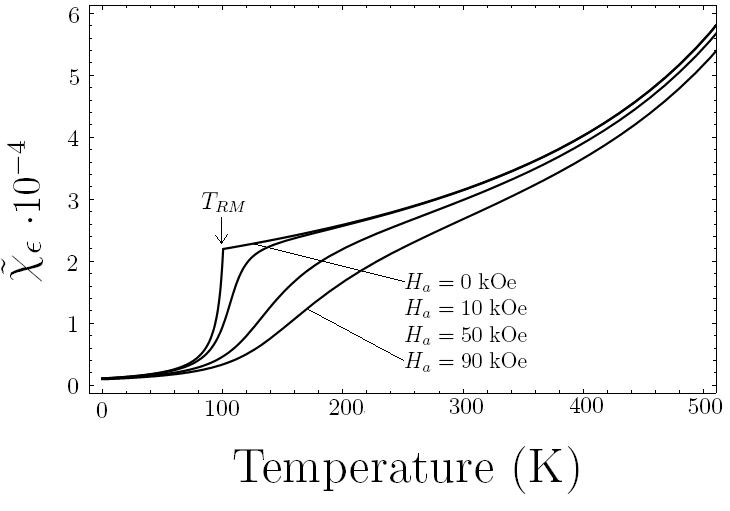}
		\caption{Dielectric susceptibility as a function of temperature for the specific values of applied magnetic field.}
	\label{fig:ep}
\end{figure}

\begin{figure}[ht]
	\centering
		\includegraphics[width=90mm]{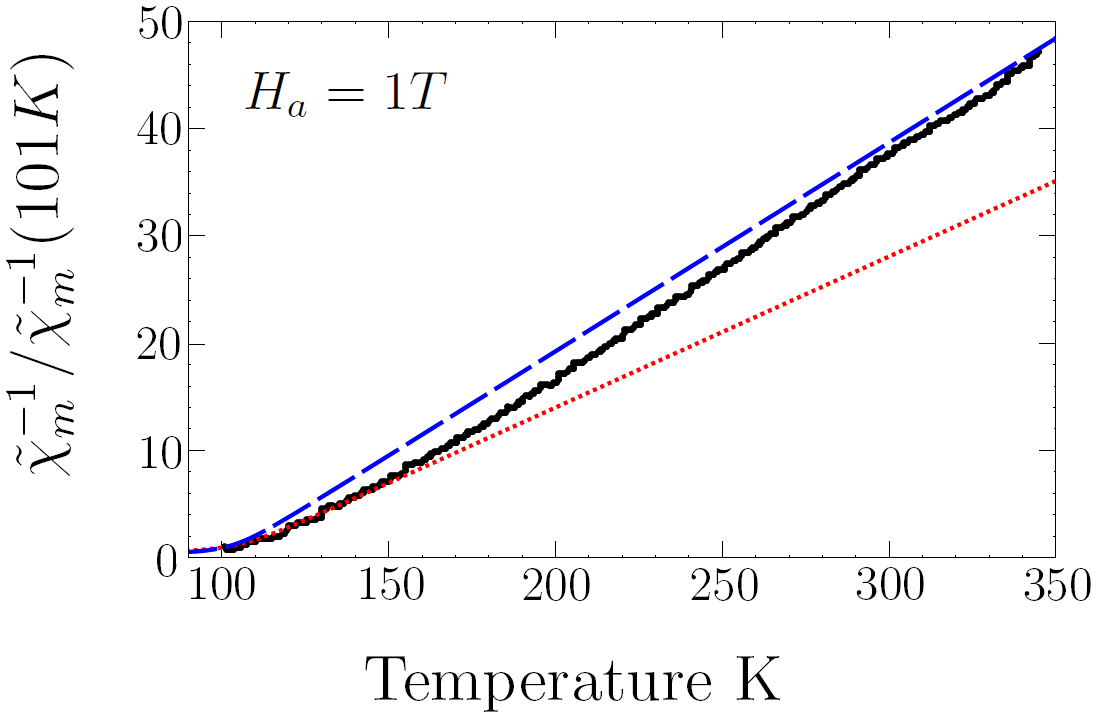}
		\caption{(Color online). Magnetic susceptibility as a function of temperature. (Solid line) The inverse molar  magnetic susceptibility of BiMnO$_3$ measured in 1T \citep{chiba} has been rescaled by its value at $T=101K$ in order to 
obtain the dimensionless quantity. Dashed and dotted lines represent the respective calculated temperature variations of the magnetic  susceptibility (\ref{chihm}); we used the same parameter values as in Fig. \ref{fig:mtn}.}
	\label{fig:chimn}
\end{figure}

In case of $\chi_{me}=0$ we calculate the inverse susceptibilities from the Landau functional as second derivative  with respect to the order parameters, i.e. $\chi_\epsilon^{-1}= \frac{\partial F^2}{\partial P^2}$, $\chi_m^{-1}= \frac{\partial F^2}{\partial M^2}$. Hence we obtain respectively:
\begin{equation}
\tilde{\chi}_\epsilon^{-1}=l\frac{T_f}{T_M}\left(\frac{T}{T_M}-\frac{T_f}{T_M}\right) + \gamma_m \frac{T_f}{T_M} \tilde{M}^2 +3l\left(\frac{T_f}{T_M}\right)^2 \tilde{P}^2,
\label{chih}
\end{equation}
\begin{equation}
\tilde{\chi}_m^{-1}= \frac{T}{T_M}-1 +\tilde{M}^2+\gamma_m\frac{T_f}{T_M} \tilde{P}^2,
\label{chihm}
\end{equation}
in which $\tilde P$ and  $\tilde M$ may be evaluated from Eqs. (\ref{p}) and (\ref{80}).
In Fig. \ref{fig:ep} and \ref{fig:chimn}  we plot the calculated temperature dependences of the susceptibilities mentioned above.  
The magnetoelectric coupling causes the enhancement of polarization around $T_{RM}$ and 
a suppression of the dielectric constant cf. Fig. \ref{fig:ep}. One  can see that at zero applied magnetic field $\tilde{\chi}_e$ decreases stepwise at $T_{RM}$  
whereas it is gradually suppressed with increasing field.
$\chi_e(T)$ exhibits a trend observed experimentally with increasing $H_a$, but the calculated
changes are too large. However, the corresponding temperature range is reproduced to much 
better accuracy than that of \citep{zhong}, where the calculated temperature $T_{RM}$ is far too low.
The magnetic-susceptibility data follow roughly the Curie-Weiss law with the paramagnetic 
Curie temperature $\Theta_{M}\approx120K$. Only the dashed curve in Fig. \ref{fig:chimn} reproduces
correctly the approximate Curie-Weiss law at high temperature. This unusual behavior from a magnetic point of view can be
understood easily from Eq. (\ref{chihm}), where the nonlinearity in magnetization (the term $\propto \tilde{M}^2$) can be dominated 
by the magnetoelectric coupling, as $T_f/T_M>>1$.

\subsection{Arrot planes}

\begin{figure}[ht]
	\centering
		\includegraphics[width=90mm]{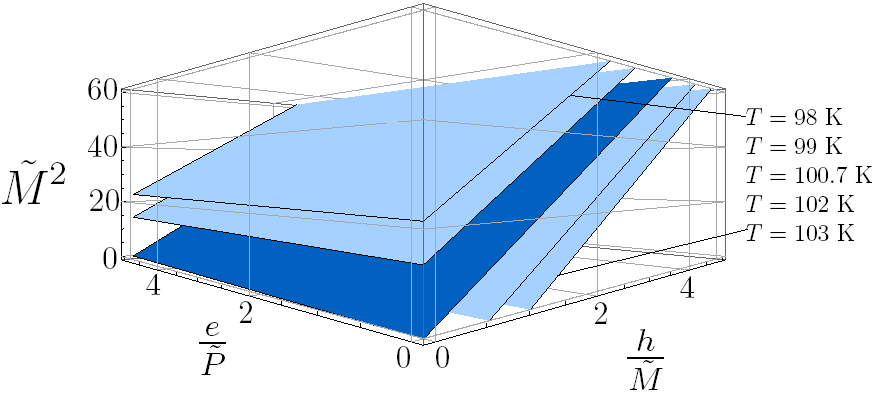}
		\caption{(Color online). Arrot set of planes for the magnetization, calculated for BiMnO$_3$ for selected temperatures. The dark plane correspond to the phase transition temperature. The parameters are
the same as in the earlier Figures.}
	\label{fig:Marrot}
\end{figure}

In systems with one order parameter a convenient way to represent their 
behavior near the phase transition temperature is to draw Arrot plots.
For a ferromagnet, the Arrot plot is a series of curves $M^2$ vs. $H_a/M$.
The same approach can be used in our case with two order parameters. The only difference
is that now instead of having a series of curves   we have sets of planes for each of the order parameters.
Namely, we have the dependencies  $M^2\left(\frac{H_a}{M}, \frac{E_a}{P}\right)$ , $P^2\left(\frac{H_a}{M}, \frac{E_a}{P}\right)$.
The plane which crosses point $(0,0)$ corresponds to the phase transition temperature in zero field. 
A representative set of the Arrot planes is drawn in Fig. \ref{fig:Marrot} for the magnetization.
One can see that the transition temperature can be determined from the 
dependence $\tilde{M}^2$ vs $e/\tilde{P}$, not only from  $\tilde{M}^2$ vs $h/\tilde{M}$! 

From (3) and (\ref{u}) after dividing both equations by the order parameter and solving  the resulting system we obtain:
\begin{widetext}
\begin{equation}
\tilde{P}^2=\frac{T_M^2}{T_f^2(l-\gamma_m^2)}\left(x_e-\gamma_m\frac{T_f}{T_M}\left(1-\frac{T}{T_M}+x_h\right)+l\frac{T_f^2}{T_M^2}\left(1-\frac{T}{T_f}\right)\right),
\label{Marrot}
\end{equation}   
\begin{equation}
\tilde{M}^2= x_h+1-\frac{T}{T_M}- \frac{\gamma_m}{l-\gamma_m^2} \frac{T_M}{T_f}\left(x_e- l\frac{T_f^2}{T_M^2}\left(\frac{T}{T_f}-1\right) - \gamma_m\frac{T_f}{T_M}\left(1-\frac{T}{T_M}+x_h\right)\right),
\end{equation} 
\end{widetext}

where $x_e= \frac{e}{\tilde{P}}$ and $x_h= \frac{h}{\tilde{M}}$.

From Eq. (\ref{Marrot}) one can also see that for $T>T_{RM}$
we still have a nonzero value of $\tilde{P}$ (cf. Eq. (\ref{MP3})). Therefore, the Arrot 
planes $P^2(H_a/M, E_a/P)$ are not important, since $T_f>T_{RM}$.  

\subsection{Specific heat}

In the framework of Ginzburg-Landau theory we may also calculate 
the specific heat as the second derivative of the free energy.
\begin{equation}
\Delta C_p= -T \left( \frac{\partial^2 F}{\partial T^2}\right)_p. 
\label{22}
\end{equation} 
Providing a similar analysis as in the previous Section we can write down the following expressions for the specific heat valid in the respecting temperature regions:

\begin{equation} 
\Delta C_p=  \left\{ \begin{array}{cccc}
\frac{l(l-2\gamma_m +1)}{2(l-\gamma_m^2)}\cdot \frac{a_0^2}{b}T  && for\ T<T_{RM}, 
\\
\frac{l}{2}\cdot \frac{a_0^2}{b}T  && for\ T>T_{RM}.
\end{array}
\right\}
\label{cv80}
\end{equation} 
 
\begin{figure}[ht]
\centering
\includegraphics[width=90mm]{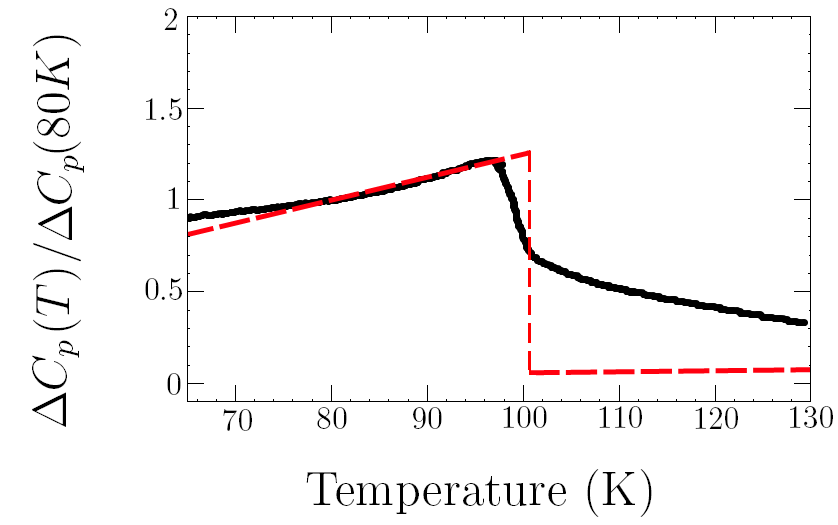}
\caption{(Color online). Temperature dependence of the specific heat for BiMnO$_3$: mean-field specific heat values - dashed line, experimental data \citep{belik} - solid line. The magnetic part of the specific
heat data  are obtained by subtracting the total
specific heat of BiScO$_3$ from that of BiMnO$_3$ \citep{belik2,belik}, since the former is not magnetic.}
\label{fig:cv2}
\end{figure}

Where $a_0=1/(T_M s_h m_0)$, $b=1/(s_h m_0^3)$.
The magnetoelectric part $\Delta C_p$ of the  specific heat calculated in this manner  is shown in Fig. \ref{fig:cv2}. The mean field values are close to the experimental data below the transition temperature $T_{RM}$. Essential differences  appear above the transition and may be
attributed  to either short-range or fluctuation effects, as discussed in the next Section.
 Nevertheless, in spite of the discrepancy in $\Delta C_p$ the Landau approach predicts the basic characteristics curve of this magnetoelectric system, surprisingly  well  at low temperatures $T<T_{RM}$.  

\section{Gaussian fluctuations of the order parameters}
 The fluctuations of the magnetization seem to have a small effect on the magnetization curve close to $T_{RM}$, but there is a discrepancy for the specific heat 
$\Delta C_p(T)$. $\Delta C_p(T)$ is taken as the difference between the total specific heat of BiMnO$_3$ and that BiScO$_3$ \citep{belik}. In this manner, $\Delta C_p(T)$ represents only the magnetic part of the specific heat.  Therefore, we discuss the role of Gaussian fluctuations of $M(\vec{r})$ and $P(\vec{r})$ on the thermodynamic properties next.
 
\subsection{Landau functional in spatially inhomogeneous case}
In  previous Sections, we made a very crucial assumption namely, that the order parameters are spatially homogeneous. The following calculations are provided in order to improve the temperature dependence of the   specific heat part $\Delta C_p(T)$  obtained above by taking into account the spatial fluctuations of the order parameters. For that purpose, we   
 introduce the effective free energy $F$ as a functional of spatially inhomogeneous order parameters \citep{strukov}:
 
\begin{equation} 
F=F_0 + \int \phi\left(M(\vec{r}),P(\vec{r}),T\right) d^3 r,
\label{DF11}     
\end{equation}
where $\phi\left(M,P,T\right)$ is the free energy functional which incorporates thermal fluctuation in equilibrium, i.e.

\begin{eqnarray}
\phi\left(M(\vec{r}),P(\vec{r}),T\right)&=& \frac{a_0(T-T_M)}{2} M(\vec{r})^2 \\\nonumber
& + & { }\frac{b}{4}M(\vec{r})^4 + \frac{c}{2}|\vec{\nabla} M(\vec{r})|^2 \\\nonumber
& + & { }\frac{\alpha_0 (T-T_f)}{2} P(\vec{r})^2+\frac{\beta}{4}P(\vec{r})^4\\\nonumber
& + & { }\frac{\delta}{2}|\vec{\nabla}P(\vec{r})|^2 + \frac{\gamma}{2}(P(\vec{r}) M(\vec{r}))^2.
\label{DF12}
\end{eqnarray}

In the present situation, we  use again dimensionless units and then the  functional $F\{M,P,T\}$ takes the form:

\begin{equation} 
F=F_0 + \frac{\Omega a_0^2 T_M^2}{b}\int_{\Omega}  \tilde{\phi}\left(\tilde{M}(\textbf{r}),\tilde{P}(\textbf{r}),T\right) d^3 \textbf{r},
\label{F2}     
\end{equation}	
where:
\begin{eqnarray}
\tilde{\phi}\left(\tilde{M}(\textbf{r}),\tilde{P}(\textbf{r}),T\right)& = & 
\frac{a_1}{2}  \tilde{P}^2+\frac{a_2}{4} \tilde{P}^4 +\frac{a_3}{2} |\vec{\nabla} \tilde{P}|^2\\\nonumber
   && { } + \frac{b_1}{2} \tilde{M}^2+\frac{1}{4}\tilde{M}^4+\frac{1}{2}|\vec{\nabla} \tilde{M}|^2\\\nonumber
   &&{ } +\frac{c_m}{2}(\tilde{P}\tilde{M})^2.
\label{DF22}
\end{eqnarray}
We use the following rescaling:
$\textbf{r}=\frac{\vec{r}}{\xi}$, $\xi=\sqrt{\frac{c}{a_0 T_M}}$,
$a_1=l \left(\frac{T_f}{T_M}\right)^2(T/T_f-1)$,
$a_2=l \left(\frac{T_f}{T_M}\right)^2$,
$a_3=n \frac{T_f}{T_M}$, $n = \frac{\delta \alpha_0 b}{c a_0 \beta}$,
$b_1=\frac{T}{T_M}-1$,
$c_m=\gamma_m \frac{T_f}{T_M}$,	and integrate over the volume $\Omega = \xi^3$ .
In doing so, we assume that the volume dependence is determined by $\xi^3$, which will be regarded 
as a fitting parameter.

\subsection{Effect of Gaussian fluctuations}
The starting point for the following calculations is to consider small fluctuations around the mean field values of the order parameters:
\begin{equation}
\tilde{M}(\textbf{r})=M_0+\delta m(\textbf{r}),
\label{m40}
\end{equation}
\begin{equation}
\tilde{P}(\textbf{r})=P_0+\delta p(\textbf{r}).
\label{p40}
\end{equation}
After substituting (\ref{m40}) and  (\ref{p40}) into (\ref{DF22}) we expand the expression
for the free energy density. For simplicity, we retain only the second order terms:
\begin{eqnarray}
\tilde{\phi}\left(\tilde{M}(\textbf{r}),\tilde{P}(\textbf{r}),T\right)& \approx & 
\frac{a_1}{2}P_0^2+\frac{a_2 }{4}P_0^4 +\frac{b_1}{2} M_0^2\\\nonumber 
& + & {} \frac{1}{4}M_0^4+\frac{c_m}{2} P_0^2M_0^2\\\nonumber 
& + & {} \delta p\left(a_1 P_0+a_2 P_0^3+c_m P_0^2 M_0\right)\\\nonumber 
& + & {} \delta m\left(b_1 M_0+M_0^3+c_m P_0 M_0^2\right)\\\nonumber 
& + & {} (\delta p)^2\left(\frac{a_1}{2}+\frac{3}{2}a_2P_0^2+\frac{1}{2}c_m M_0^2\right)\\\nonumber 
& + & {} (\delta  m)^2\left(\frac{b_1}{2}+\frac{3}{2}M_0^2+\frac{1}{2}c_m P_0^2\right)\\\nonumber 
& + & {} \frac{c_m}{2}P_0M_0 \delta m\delta p \\\nonumber 
& + & {} \frac{1}{2} |\vec{\nabla}\delta m|^2+\frac{a_3}{2} |\vec{\nabla}\delta p|^2.
\label{gF}
\end{eqnarray}

The constant expression:
\begin{equation}
\phi_0=\frac{a_1}{2}P_0^2+\frac{a_2 }{4}P_0^4+
\frac{b_1}{2} M_0^2+\frac{1}{4}M_0^4+\frac{c_m}{2} P_0^2M_0^2,
\label{F0}
\end{equation}

gives the mean filed value of the free energy density. Linear terms in $\{\delta m,\delta p\}$ vanish,
because the mean field solution $\{M_0,P_0\}$ minimizes  the free energy $F$.
In effect, we obtain the contribution to the free energy coming from the fluctuations of the order parameters in the form:
\begin{widetext}  
\begin{equation} 
\delta F= C_m\int\left\{ A_m(\delta p)^2 +B_m(\delta m)^2 +\frac{1}{2} |\vec{\nabla}\delta m|^2+\frac{a_3}{2} |\vec{\nabla}\delta p|^2 + \gamma_{me} \delta m\delta p\right\}d^3 \textbf{r},
\label{ges}     
\end{equation}
\end{widetext}
where: $A_m=\frac{a_1}{2}+\frac{3}{2}a_2 P_0^2+\frac{1}{2}c_m M_0^2$,
$B_m=\frac{b_1}{2}+\frac{3}{2}M_0^2+\frac{1}{2}c_m P_0^2$,
$\gamma_{me}=\frac{c_m}{2} P_0 M_0$,
and $C_m=\frac{\Omega a_0^2 T_M^2}{b}$.

With the help of the expression for the $\delta F$ obtained above we calculate in the Appendix A 
the explicit form of the partition function $\mathcal{Z}$, which takes into account spatial fluctuations of
the two interacting order parameters.

\subsection{Specific heat} 
To calculate explicitly the contribution of the fluctuations to the specific heat we use the partition function (\ref{7}). The free energy  part due to fluctuations can be written as:
\begin{widetext}
\begin{eqnarray} 
\delta F&=&-k_B T \ln \mathcal{Z}\\\nonumber 
&=& {}-\frac{k_B T}{2} \sum_{\textbf{\textbf{k}}} \ln\left\{\frac{\pi^2}{4}\left(\frac{k_B T}{C_m}\right)^2\frac{1}{|(A_m+\frac{1}{2}a_3 \textbf{k}^2)(B_m+\frac{1}{2}\textbf{k}^2)+\frac{1}{4}\gamma_{me}^2|}\right\}.
\end{eqnarray}  
\end{widetext}
We change the summation over $\textbf{k}$ into integration and have:
\begin{widetext}  
\begin{equation}
\delta F= -\frac{k_B T}{2} \int_0^{k_{max}} \frac{d^3 \textbf{k}}{(2\pi)^3} 
 \ln\left\{\frac{\pi^2}{4}\left(\frac{k_B T}{C_m}\right)^2\frac{1}{|(A_m+\frac{1}{2}a_3 \textbf{k}^2)(B_m+\frac{1}{2}\textbf{k}^2)+\frac{1}{4}\gamma_{me}^2|}\right\}.
\label{100}
\end{equation}
\end{widetext}

\begin{figure}[ht]
\centering
\includegraphics[width=90mm]{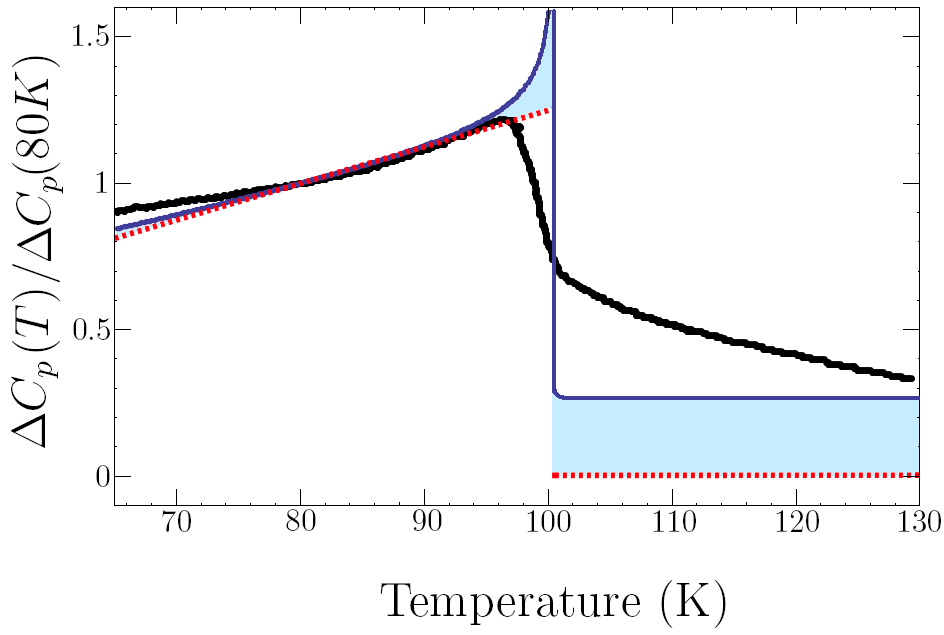}
\caption{(Color online). Temperature dependence of the specific heat for BiMnO$_3$. Dotted line: the mean-field part of the  specific heat; thick solid line: experimental data \citep{belik} was taken as $C_p| _{BiMnO_3}-C_p| _{BiScO_3}$. Thin solid line: specific heat after taking into account the Gaussian thermal fluctuations calculated for  $\xi=2.71 \cdot 10^{-8}[m]$ and $n=1$. }
\label{fig:cv3}
\end{figure}

After differentiating twice (\ref{100}) (cf. Eq. (\ref{22})), we obtain the part $\delta C_p$ for the specific heat including Gaussian fluctuations. 
Finally, the total specific heat becomes
\begin{equation}
\Delta C_p=-T \frac{\partial^2\phi_0}{\partial T ^2}\cdot \frac{a_0^2  \Omega}{b}+\delta C_p,
\label{cv}
\end{equation} 
where we have added  $\delta C_p$ to the mean-field part.
Following our previous notation we set the integration limit as $k_{max}=\xi \frac{\pi}{a}$, where we take
the lattice parameter $a=9.5415 \AA $ \citep{belik}.
In Fig. \ref{fig:cv3} we compare the theoretical results for the specific heat with the temperature dependence of  $\Delta C_p = C_p| _{BiMnO_3}-C_p| _{BiScO_3}$. We see that the fluctuations overestimate the experimental behavior  for $T<T_{RM}$ and underestimate the data for $T>T_{RM}$. We attribute this (cf. Sec. III B) to  the role of short-range order which gradually disappears  as T increases above $T_{RM}$. This is also the reason why the mean-field results match the experiment well for $T<T_{RM}$.   
 
\subsection{Correlation lengths in multiferroics}
In an analogous manner one can calculate the  evolution  of the correlation length through the magnetic phase transition.

The correlation function of an order parameter $\psi(\vec{r})$ in two distant points
is defined as:
\begin{equation}
g(\vec{r},\vec{r'})=\left\langle (\psi(\vec{r})-\bar{\psi})(\psi(\vec{r'})-\bar{\psi})\right\rangle=\left\langle \delta \psi(\vec{r})\delta \psi(\vec{r'})\right\rangle.
\label{8}
\end{equation}

After a  Fourier transform  we can write $g(\vec{r},\vec{r'})$ as:
\begin{equation} 
\left\langle \delta \psi(\vec{r_1})\delta \psi(\vec{r_2})\right\rangle  =  
 \sum_{k_1, k_2}\left\langle \psi_{k_1}^* \psi_{k_2} e^{i (\vec{k_2}-\vec{k_1})\cdot \vec{r_1}}e^{i \vec{k_2}\cdot (\vec{r_2}-\vec{r_1})}   \right\rangle
\label{9}
\end{equation} 
 
and finally:
 \begin{eqnarray} 
\left\langle \delta \psi(\vec{r_1})\delta \psi(\vec{r_2})\right\rangle= \sum_{k}\left\langle \delta \psi_{-k} \delta \psi_k \right\rangle e^{i \vec{k}\cdot \vec{r}},
\label{10} 
\end{eqnarray} 
where $\vec{r}=\vec{r_2}-\vec{r_1}$. 
 
To evaluate the coherence length for both the magnetic and the electric order parameters we 
need to calculate $\left\langle \delta m_{\textbf{\textbf{-k}}} \delta m_{\textbf{\textbf{k}}} \right\rangle$ and  $\left\langle \delta p_{\textbf{\textbf{-k}}} \delta p_{\textbf{\textbf{k}}} \right\rangle$. Using (\ref{mk4}) and (\ref{pk4}) we obtain the following expressions:
\begin{equation}
\left\langle \delta m_{\textbf{\textbf{-k}}} \delta m_{\textbf{\textbf{k}}} \right\rangle=\left\langle \delta m_{\textbf{\textbf{1k}}}^2 + \delta m_{\textbf{\textbf{2k}}}^2 \right\rangle,
\label{11}
\end{equation} 
\begin{equation}
\left\langle \delta p_{\textbf{\textbf{-k}}} \delta p_{\textbf{\textbf{k}}} \right\rangle=\left\langle \delta p_{\textbf{\textbf{1k}}}^2 + \delta p_{\textbf{\textbf{2k}}}^2 \right\rangle.
\label{12}
\end{equation} 
\begin{equation}
\frac{1}{2}\left\langle \delta p_{\textbf{\textbf{-k}}} \delta m_{\textbf{\textbf{k}}} +\delta p_{\textbf{\textbf{k}}} \delta m_{\textbf{\textbf{-k}}} \right\rangle=\left\langle \delta p_{\textbf{\textbf{1k}}}m_{\textbf{\textbf{1k}}} + \delta p_{\textbf{\textbf{2k}}}m_{\textbf{\textbf{2k}}} \right\rangle.
\label{mkpk}
\end{equation}    
The average  values $\left\langle \delta m_{\textbf{\textbf{-k}}} \delta m_{\textbf{\textbf{k}}} \right\rangle$ and $\left\langle \delta p_{\textbf{\textbf{-k}}} \delta p_{\textbf{\textbf{k}}} \right\rangle$ 
can be calculated using the matrix $\textbf{A}_{\textbf{k}}$ defined in the Appendix A:
\begin{equation}
\left\langle \delta m_{\textbf{\textbf{-k}}} \delta m_{\textbf{\textbf{k}}} \right\rangle =\frac{\int \mathcal{D}\eta_{\textbf{k}} e^{-\beta \sum_{\textbf{k}} \vec{\eta}^{T}_{\textbf{k}} \textbf{A}_{\textbf{k}} \vec{\eta}^{T}_{\textbf{k}}} (\delta m_{1\textbf{k}}^2 +\delta m_{2\textbf{k}}^2)}{\int \mathcal{D}\eta_{\textbf{k}} e^{-\beta \sum_{\textbf{k}} \vec{\eta}^{T}_{\textbf{k}} \textbf{A}_{\textbf{k}} \vec{\eta}^{T}_{\textbf{k}}}},
\label{13}
\end{equation}
 \begin{equation}
\left\langle \delta p_{\textbf{\textbf{-k}}} \delta p_{\textbf{\textbf{k}}} \right\rangle =\frac{\int \mathcal{D}\eta_{\textbf{k}} e^{-\beta \sum_{\textbf{k}} \vec{\eta}^{T}_{\textbf{k}} \textbf{A}_{\textbf{k}} \vec{\eta}^{T}_{\textbf{k}}} (\delta p_{1\textbf{k}}^2 +\delta p_{2\textbf{k}}^2)}{\int \mathcal{D}\eta_{\textbf{k}} e^{-\beta \sum_{\textbf{k}} \vec{\eta}^{T}_{\textbf{k}} \textbf{A}_{\textbf{k}} \vec{\eta}^{T}_{\textbf{k}}}}.
\label{14}
\end{equation}  

If $\textbf{A}_{\textbf{k}}$ is symmetric for real Gaussian integrals as in (\ref{13}) and (\ref{14}) we have:
 \begin{equation}
\int d\vec{\eta}_{\textbf{k}}  e^{\frac{1}{2}\vec{\eta}_{\textbf{k}}^T\textbf{A}_{\textbf{k}}\vec{\eta}_{\textbf{k}}}\eta_i\eta_j=(2\pi)^{\frac{N}{2}}(det \textbf{A})^{-\frac{1}{2}}A_{ij}^{-1},
\label{RR}
\end{equation} 
where $A_{ij}$ - is an element of the $\textbf{A}_{\textbf{k}}$ matrix in the $i$th - row and $j$th-column,
$\eta_i$ for $i=1,2,3,4$ is an element of the vector $\vec{\eta}_{\textbf{k}}$.

We obtain following expressions:
\begin{equation}
\left\langle \delta p_{\textbf{\textbf{-k}}} \delta p_{\textbf{\textbf{k}}} \right\rangle =A_{11}^{-1}+A_{22}^{-1}= \frac{k_B T}{2 C_m(A_m+\frac{1}{2}a_3 \textbf{k}^2)},
\label{15}
\end{equation} 

 \begin{equation}
\left\langle \delta m_{\textbf{\textbf{-k}}} \delta m_{\textbf{\textbf{k}}} \right\rangle =A_{33}^{-1}+A_{44}^{-1}= \frac{k_B T}{2 C_m(B_m+\frac{1}{2}\textbf{k}^2)},
\label{16}
\end{equation} 
and for the cross-correlations we have:
\begin{equation}
\left\langle \delta p_{\textbf{\textbf{1k}}}m_{\textbf{\textbf{1k}}} + \delta p_{\textbf{\textbf{2k}}}m_{\textbf{\textbf{2k}}} \right\rangle=A_{31}^{-1}+A_{42}^{-1}=\frac{k_B T}{C_m\gamma_{me}}. 
\end{equation}
 
Finally, the correlation functions for the magnetic and electric subsystems take the form:

\begin{eqnarray} 
g_p(\vec{r_1},\vec{r_2}) &=&\left\langle \delta p(\vec{r_1}) \delta p(\vec{r_2}) \right\rangle \\\nonumber  &=&{} \int{\frac{d^3 \textbf{k}}{(2\pi)^3}}\left\langle \delta p_{\textbf{\textbf{-k}}} \delta p_{\textbf{\textbf{k}}} \right\rangle e^{i \textbf{\textbf{k}}\cdot (\vec{r_2}-\vec{r_1})}\\\nonumber 
&=& {} \int{\frac{d^3 \textbf{k}}{(2\pi)^3}}\cdot \frac{k_B T}{2 C_m(A_m+\frac{1}{2}a_3 \textbf{k}^2)}e^{i \textbf{\textbf{k}}\cdot (\vec{r_2}-\vec{r_1})},
\label{17}
\end{eqnarray} 

\begin{eqnarray} 
g_m(\vec{r_1},\vec{r_2})&=&\left\langle \delta m(\vec{r_1}) \delta m(\vec{r_2}) \right\rangle \\\nonumber &=&{} \int{\frac{d^3 \textbf{k}}{(2\pi)^3}}\left\langle \delta m_{\textbf{\textbf{-k}}} \delta m_{\textbf{\textbf{k}}} \right\rangle e^{i \textbf{\textbf{k}}\cdot (\vec{r_2}-\vec{r_1})}\\\nonumber 
&=& {} \int{\frac{d^3 \textbf{k}}{(2\pi)^3}}\cdot \frac{k_B T}{2 C_m(B_m+\frac{1}{2} \textbf{k}^2)}e^{i \textbf{\textbf{k}}\cdot (\vec{r_2}-\vec{r_1})}.
\label{18}
\end{eqnarray}
Using:   
\begin{equation*}         
\int{\frac{d^3 k}{(2\pi)^3}} \frac{1}{k^2+a^2} e^{i \vec{k}\cdot\vec{r}}=\frac{e^{-ar}}{4\pi r},
\end{equation*}           
we obtain the correlation function in the Ornstein-Zernike form: 

\begin{equation}
g_p(\vec{r_1},\vec{r_2})=\frac{k_B T}{C_m a_3}\cdot\frac{e^{-\sqrt{\frac{2A_m}{a_3}}|\vec{r_2}-\vec{r_1}|}}{4\pi |\vec{r_2}-\vec{r_1}|}\equiv \frac{k_B T}{C_m a_3}\cdot\frac{e^{-\frac{|\vec{r_2}-\vec{r_1}|}{\xi_m}}}{4\pi |\vec{r_2}-\vec{r_1}|},
\label{19}
\end{equation}
 
\begin{equation}
g_m(\vec{r_1},\vec{r_2})=\frac{k_B T}{C_m}\cdot\frac{e^{-\sqrt{2B_m}|\vec{r_2}-\vec{r_1}|}}{4\pi |\vec{r_2}-\vec{r_1}|}\equiv \frac{k_B T}{C_m}\cdot\frac{e^{-\frac{|\vec{r_2}-\vec{r_1}|}{\xi_p}}}{4\pi |\vec{r_2}-\vec{r_1}|},
\label{20}
\end{equation}
with the correlation lengths:
\begin{equation}
\xi_p =\sqrt{\frac{a_3}{2A_m}},\ \xi_m =\sqrt{\frac{1}{2B_m}}.
\label{21}
\end{equation}
 
From Eqs. (\ref{19}) and (\ref{20}) we see that when the coherence length is close to zero the correlation  function  becomes equal to zero as well. On the other hand when the coherence length is large the correlation  function decreases as $\propto \frac{1}{|\vec{r_2}-\vec{r_1}|}$, The correlation radii in this region are significantly greater than the lattice constant.

\begin{figure}[ht]
\centering
\includegraphics[width=90mm]{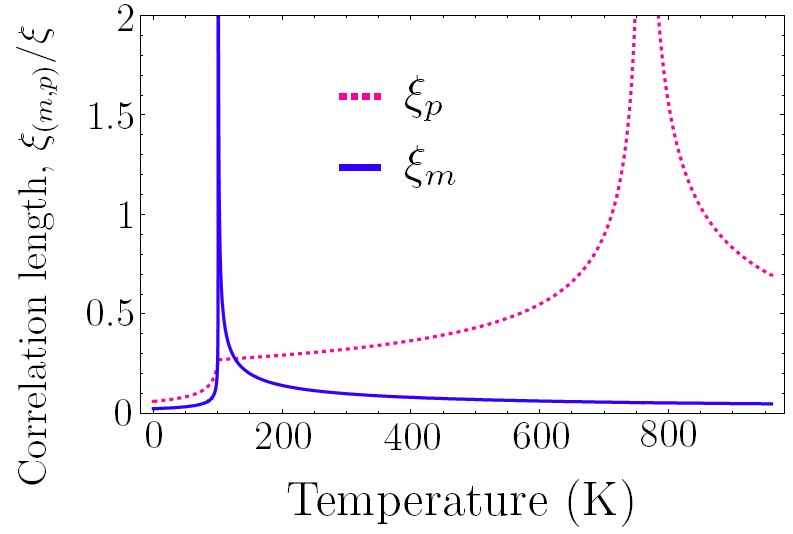}
\caption{(Color online). Temperature dependence of the coherence lengths for magnetic (solid line) and electric (dotted line) subsystems  calculated for BiMnO$_3$. The parameters we use are the same as in the earlier Figures.}
\label{fig:kh}
\end{figure}

We can see that close to ferroelectromagnetic phase transition temperature both coherence 
lengths  behave in a similar way. Hence, we assume that below $T_{RM}$, where both magnetic and 
electric ordering is observed, the coherence length coalesces as $T$ decreases ($T\rightarrow 0$) (cf. Fig. \ref{fig:kh}).
It can also be seen (from the Fig. \ref{fig:kh}) that the order parameter corresponding 
to the lower critical temperature experience smaller fluctuations near the phase transition.

We calculate the cross correlations of the order parameters, which take the form: 
\begin{eqnarray}
g_{pm}(\vec{r_1},\vec{r_2})&=& \left\langle \delta m(\vec{r_1}) \delta p(\vec{r_2}) \right\rangle\\\nonumber 
&=& {} \sum_\textbf{k} \left\langle \delta m_{\textbf{1k}} \delta p_{\textbf{1k}}+\delta m_{\textbf{2k}} \delta p_{\textbf{2k}} \right\rangle e^{i\vec{\textbf{k}}\cdot \vec{\textbf{r}}}\\\nonumber
&=& {} \frac{1}{2 \pi^2} \cdot \frac{k_B T}{C_m\gamma_{me} r}\int_0^{k_{max}} k \sin(k r)dk\\\nonumber
&=& {} \frac{k_B T}{ 2 \pi^2 C_m\gamma_{me} r^3}(\sin(k_{max} r)-r k_{max} \cos(k_{max} r)).
\label{gpm}
\end{eqnarray}

We see that the correlation length $\xi_p$ near and above the magnetic phase transition is enhanced,
whereas the cross correlations exhibit an oscillatory behavior and evolve continuously through $T_{RM}$.     
\section{Conclusions}

In this paper we examined a simple Landau approach for a system with two single-component order
parameters representing ferroelectricity and ferromagnetism, respectively. The obtained results are  consistent with experiment.      
Magnetoelectric coupling introduces a strong renormalization of the ferromagnetic transition temperature
(enhancing it by a factor of $50$) causing the magnetic phase transition to be observed at $T_{RM}=100.5 K$. A number of  coupling effects such as the enhancement of electric polarization, the anomaly in the dielectric susceptibility, and a fairly large negative magnetocapacitance, occur concomitantly  in the vicinity of $T_{RM}$. We introduced a simple extension of the Arrot plot which we called in text  \textit{Arrot planes}. We consider this concept to be potential useful in evaluation of the phase transition temperature while using only field dependence of unbounded with it order parameter. We have also estimated the contribution of the Gaussian fluctuations to  the specific heat and have noted that short-range-order effects are not accounted properly.  It would be important to understand the microscopic reasons of such a strong enhancement of the magnetic exchange interactions caused by a  monoclinic lattice distortion which leads to the appearance of the ferroelectric dipole moments. 
 
\begin{acknowledgments}
 We would like to cordially thank Leszek J. Spalek from the Cavendish Laboratory, Cambridge, for suggesting the problem
 and numerous discussions, as well as for his critical reading of the manuscript. 
The work was supported by the Grant No. NN 202 128 736 from Ministry of Science and Higher Education.
\end{acknowledgments} 

\appendix
\section{Partition function with inclusion of Gaussian fluctuations}

If we assume that the fluctuations do not change rapidly in space then we can estimate their local value in terms of their Fourier components:
\begin{equation} 
\delta p(\textbf{r})=\sum_{\textbf{\textbf{k}}}\delta p_{\textbf{\textbf{k}}} e^{i \textbf{\textbf{k}} \cdot \textbf{\textbf{r}}},
\label{furjerp}     
\end{equation}	          
\begin{equation} 
\delta m(\textbf{r})=\sum_{\textbf{\textbf{k}}}\delta m_{\textbf{\textbf{k}}} e^{i\textbf{\textbf{k}}\cdot\textbf{\textbf{r}}},
\label{furjerm}     
\end{equation}

where $\textbf{k}=\xi \vec{k}$.
We also assume that $e^{i\vec{k}\cdot\vec{r}}$ satisfies periodic boundary conditions.

After substituting (\ref{furjerp}) and (\ref{furjerm}) into (\ref{ges}):

\begin{eqnarray} 
\delta F&=& C_m \gamma_{me}\sum_{\textbf{\textbf{k}}}\sum_{\textbf{\textbf{q}}}\delta m_{\textbf{\textbf{k}}}\delta p_{\textbf{\textbf{q}}} \int d^3 \textbf{r} e^{i(\textbf{\textbf{k}}+\textbf{\textbf{q}})\cdot \textbf{\textbf{r}}} \\\nonumber  
& + & { } C_m\sum_{\textbf{\textbf{k}}}\sum_{\textbf{\textbf{q}}}\left[B_m-\frac{1}{2} \textbf{k}\cdot\textbf{q}\right]\delta m_{\textbf{\textbf{k}}}\delta m_{\textbf{\textbf{q}}} \int d^3 \textbf{r} e^{i(\textbf{\textbf{k}}+\textbf{\textbf{q}})\cdot \textbf{\textbf{r}}}\\\nonumber 
& + & {} C_m\sum_{\textbf{\textbf{k}}}\sum_{\textbf{\textbf{q}}}\left[A_m-\frac{1}{2} a_3 \textbf{k}\cdot\textbf{q}\right]\delta p_{\textbf{\textbf{k}}}\delta p_{\textbf{\textbf{q}}} \int d^3 \textbf{r} e^{i(\textbf{\textbf{k}}+\textbf{\textbf{q}})\cdot \textbf{\textbf{r}}}.
\label{sumaF}     
\end{eqnarray}
Taking into consideration that $\int{d^3 r e^{i(\textbf{\textbf{k}}+\textbf{\textbf{q}})\cdot \textbf{\textbf{r}}}}=\delta_{\textbf{\textbf{k}},\textbf{\textbf{-q}}}$
and after rewriting the last term as:
\begin{eqnarray} 
\sum_{\textbf{\textbf{k}}}\sum_{\textbf{\textbf{q}}}\delta m_{\textbf{\textbf{k}}}\delta p_{\textbf{\textbf{q}}}\delta_{\textbf{\textbf{k}},\textbf{\textbf{-q}}}&=& {}\frac{1}{2}\left[\sum_{\textbf{\textbf{k}}}\delta m_{\textbf{\textbf{k}}}\delta p_{\textbf{\textbf{-k}}} +\sum_{\textbf{\textbf{q}}}\delta m_{\textbf{\textbf{-q}}}\delta p_{\textbf{\textbf{q}}}\right]\\\nonumber
&=&{}\frac{1}{2}\sum_{\textbf{\textbf{k}}}\left[\delta m_{\textbf{\textbf{k}}}\delta p_{\textbf{\textbf{-k}}}+\delta m_{\textbf{\textbf{-k}}}\delta p_{\textbf{\textbf{k}}}\right],
\label{sumaF2}     
\end{eqnarray}
we obtain the following expression for the free energy:
\begin{eqnarray} 
\delta F&=& C_m\sum_{\textbf{\textbf{k}}}(A_m+\frac{1}{2} a_3 \textbf{k}^2)\delta p_{\textbf{\textbf{k}}}\delta p_{\textbf{\textbf{-k}}} \\\nonumber 
& + & {} C_m\sum_{\textbf{\textbf{k}}}(B_m+\frac{1}{2}\textbf{k}^2)\delta m_{\textbf{\textbf{k}}}\delta m_{{\textbf{\textbf{-k}}}}\\\nonumber 
& + & {}\frac{1}{2}C_m\gamma_{me}\sum_{\textbf{\textbf{k}}}\left[\delta m_{\textbf{\textbf{k}}}\delta p_{\textbf{\textbf{-k}}}+\delta m_{\textbf{\textbf{-k}}}\delta p_{\textbf{\textbf{k}}}\right].
\label{sumaF3}     
\end{eqnarray}

In order to evaluate the value of the fluctuations we have to take an average of all the possible 
configurations. One can define the statistical sum for a system with two order parameters 
as an integral over all existing profiles for each of the order parameters:
\begin{equation}
\mathcal{Z}=\prod_{\textbf{\textbf{k}}}\int{\mathcal{D}\left(\delta m_{\textbf{\textbf{k}}}\right) \mathcal{D}\left(\delta p_{\textbf{\textbf{k}}}\right) e^{-\frac{\delta F(\delta m, \delta p)}{k_B T}}}.
\label{z2}     
\end{equation}

Because $\delta m_{\textbf{\textbf{k}}}$ and  $\delta p_{\textbf{\textbf{k}}}$ 
are complex numbers we can represent them in the following way:
\begin{equation}
\delta m_{\textbf{\textbf{k}}}=\delta m_{\textbf{\textbf{1k}}}+i\cdot \delta m_{\textbf{\textbf{2k}}},
\label{mk4}    
\end{equation}
\begin{equation}
\delta p_{\textbf{\textbf{k}}}=\delta p_{\textbf{\textbf{1k}}}+i \cdot\delta p_{\textbf{\textbf{2k}}}. 
\label{pk4}
\end{equation}
As may be seen $\delta m_{\textbf{\textbf{k}}}$ and $\delta m_{\textbf{\textbf{-k}}}$ are not independent. 
In order to avoid double counting  for $\delta m_{\textbf{\textbf{k}}}$ and $\delta m_{\textbf{\textbf{-k}}}$  we have to
take only wave vectors $\textbf{k}$  with $k_z>0$
(the same applies for $\delta p$).
Thus, the partition function representing the Gaussian fluctuations takes the form: 
\begin{widetext}
\begin{equation}
\mathcal{Z}=\prod_{\textbf{\textbf{k}}, k_z>0}\int_{-\infty}^\infty \mathcal{D}\left(\delta m_{\textbf{\textbf{1k}}}\right)\int_{-\infty}^\infty \mathcal{D}\left(\delta m_{\textbf{\textbf{2k}}}\right) \int_{-\infty}^\infty \mathcal{D}\left(\delta p_{\textbf{\textbf{1k}}}\right) \int_{-\infty}^\infty \mathcal{D}\left(\delta p_{\textbf{\textbf{2k}}}\right) e^{-\frac{\delta F(\delta m, \delta p)}{k_B T}},
\label{1}     
\end{equation}          
\end{widetext}
where the corresponding free energy functional is
\begin{eqnarray}
\delta F&=&2C_m\sum_{\textbf{\textbf{k}},k_z>0}(A_m+\frac{1}{2} a_3 \textbf{k}^2)(\delta  p^2_{\textbf{\textbf{1k}}}+\delta  p^2_{\textbf{\textbf{2k}}})\\\nonumber 
& + &{} 2C_m\sum_{\textbf{\textbf{k}},k_z>0}(B_m+\frac{1}{2}\textbf{k}^2)(\delta  m^2_{\textbf{\textbf{1k}}}+\delta  m^2_{\textbf{\textbf{2k}}})\\\nonumber 
& + &{} 2C_m\gamma_{me}\sum_{\textbf{\textbf{k}},k_z>0}(\delta  m_{\textbf{\textbf{1k}}}\delta  p_{\textbf{\textbf{1k}}}+\delta  m_{\textbf{\textbf{2k}}}\delta  p_{\textbf{\textbf{2k}}}). 
\label{FFF}     
\end{eqnarray}
For further calculations we use matrix notation: 

\begin{equation}
\vec{\eta}_{\textbf{k}}= 
\begin{pmatrix} 
\delta  p_{\textbf{\textbf{1k}}} \\
\delta  p_{\textbf{\textbf{2k}}}  \\
\delta  m_{\textbf{\textbf{1k}}}  \\
\delta  m_{\textbf{\textbf{2k}}}\\
\end{pmatrix}, 
\end{equation}
\begin{equation}
\vec{\eta}^{T}_{\textbf{k}}=(\delta  p_{\textbf{\textbf{1k}}}, \delta  p_{\textbf{\textbf{2k}}}, \delta  m_{\textbf{\textbf{1k}}}, \delta  m_{\textbf{\textbf{2k}}}),
\end{equation}
 and
\begin{widetext} 
\begin{equation}
\textbf{A}_{\textbf{k}}= \begin{pmatrix}
4C_m(A_m+\frac{1}{2}a_3 \textbf{k}^2) & 0 & 2C_m\gamma_{me}&0 \\
0 & 4C_m (A_m+\frac{1}{2}a_3 \textbf{k}^2) & 0 & 2C_m \gamma_{me} \\
2C_m \gamma_{me}&0 & 4C_m(B_m+\frac{1}{2}\textbf{k}^2) & 0\\
0 & 2C_m\gamma_{me} & 0 & 4C_m(B_m+\frac{1}{2}\textbf{k}^2)
\end{pmatrix}. 
\end{equation}
\end{widetext}

In these terms the statistical sum $\mathcal{Z}$  can be written as:
\begin{equation}
\mathcal{Z}=\int \mathcal{D}\eta_{\textbf{k}} e^{-\beta \sum_{\textbf{k}} \vec{\eta}^{T}_{\textbf{k}} \textbf{A}_{\textbf{k}} \vec{\eta}^{T}_{\textbf{k}}}\equiv\prod_{\textbf{k}}\mathcal{Z}_{\textbf{k}}. 
\label{ZKK}
\end{equation}
Because the matrix $\textbf{A}_{\textbf{k}}$ is symmetric and the vector $\vec{\eta}_{\textbf{k}}$ is real, we can use the expression for the last Gaussian integral:  
 \begin{equation}
\mathcal{Z}_{\textbf{k}}=(2\pi)^{\frac{D}{2}}(det \textbf{A}_{\textbf{k}})^{-\frac{1}{2}},
\label{ZKG}
\end{equation}
where D is the dimension of the vector $\vec{\eta}_{\textbf{k}}$ which in our case is equal to 4.
The final expression for $\mathcal{Z}_{\textbf{k}}$ after  diagonalization of (\ref{FFF}) can be written as:
\begin{equation}
\mathcal{Z}_{\textbf{\textbf{k}}}=\frac{\pi^2}{4}\left(\frac{k_B T}{C_m}\right)^2\frac{1}{|(A_m+\frac{1}{2}a_3 \textbf{k}^2)(B_m+\frac{1}{2} \textbf{k}^2)-\frac{1}{4}\gamma_{me}^2|}.
\label{rozk}
\end{equation}   
The corresponding total statistical sum is:
\begin{equation}
\mathcal{Z}=\prod_{\textbf{\textbf{k}}, k_z>0} \mathcal{Z}_{\textbf{\textbf{k}}}
\label{7}
\end{equation}
This expression is used in Sec. IVC to calculate the specific heat. 

\section{Angular degrees of freedom in order parameter fluctuations } 
\begin{wrapfigure}{l}{0.25\textwidth}
  \begin{center}
    \includegraphics[width=0.20\textwidth]{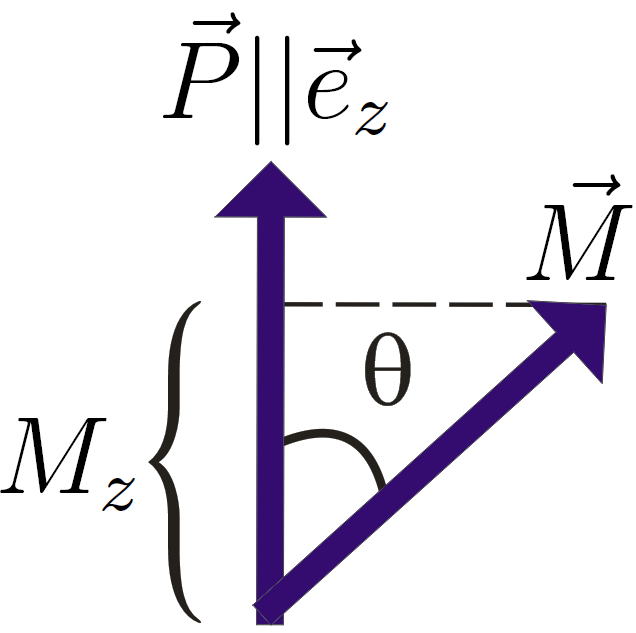}
    
  \end{center}
  \caption{(Color online). Model picture of the order parameters arrangement.}
  \label{fig:mzp}
\end{wrapfigure}

In our calculations we see a big discrepancy in specific heat near the phase transition $T_M$.
Here we present a brief discussion of possible way to improve accordance to the experimental data by considering
a little bit more realistic model. We assume that our order parameters are three dimensional. Due to the fact that our main interest lies in region near ferromagnetic phase transition and because $T_f >> T_M$ we can consider a following simplification: $\vec{P}(\vec{r})=P(\vec{r})\vec{e}_z$.
The main idea of proposed improvement is to take into account the angular fluctuation of $\vec{M}$ near the phase transition. 
If we apply the orientation of the order parameters as presented at Fig. \ref{fig:mzp} we obtain following Landau free energy potential

\begin{equation}
\Delta F = \frac{a}{2}M_z^2 \cos^2\Theta + \frac{b}{4}M_z^4 \cos^4\Theta + \frac{\alpha}{2}P^2 + \frac{\beta}{4}P^4  + \gamma(PM_z)^2\cos^2\Theta.
\label{B1}
\end{equation} 

In the spirit of mode-mode coupling approach we can present the following two terms from \ref{B1} as:
\begin{equation}
(M_z\cos(\Theta))^2\approx \left\langle M_z^2\right\rangle \cos^2\Theta + \left\langle \cos^2\Theta\right\rangle M_z^2 -\left\langle M_z^2\right\rangle \left\langle \cos^2\Theta \right\rangle 
\label{B2}
\end{equation} 
and  
\begin{equation}
(M_z\cos(\Theta))^4\approx \left\langle M_z^2\right\rangle \left\langle \cos^2\Theta\right\rangle\left\{ M_z^2 + \left\langle M_z^2\right\rangle \cos^2\Theta - \left\langle \cos^2\Theta\right\rangle \left\langle M_z^2\right\rangle \right\}.
\label{B3}
\end{equation}

After substitution \ref{B2} and \ref{B3} in \ref{B1} the free energy potential will have the form:

\begin{eqnarray}
 \Delta F&=& M_z^2 \left\{\frac{a}{2}\left\langle \cos^2\Theta\right\rangle + \frac{b}{4}\left\langle M_z^2\right\rangle \left\langle \cos^2\Theta\right\rangle + \frac{\gamma}{2} P^2 \left\langle \cos^2\Theta \right\rangle\right\} \\ \nonumber
  &&{} + \cos^2\Theta \left\{\frac{a}{2}\left\langle M_z^2\right\rangle + \frac{b}{4}\left\langle M_z^2\right\rangle^2 \left\langle \cos^2\Theta\right\rangle + \frac{\gamma}{2} P^2 \left\langle M_z^2\right\rangle\right\}\\\nonumber 
  &&{} + P^2 \left\{\frac{\alpha}{2} + \frac{\beta}{4}\left\langle P^2\right\rangle  + \frac{\gamma}{2} \left\langle M_z^2 \right\rangle\left\langle \cos^2\Theta \right\rangle\right\}\\\nonumber 
  &&{} -\left\langle M_z^2 \right\rangle\left\langle \cos^2\Theta \right\rangle\left\{\frac{a}{2}P^2 + \frac{b}{4}\left\langle M_z^2 \right\rangle\left\langle \cos^2\Theta \right\rangle\right\}
\label{DDF}
\end{eqnarray}


\begin{thebibliography}{26}

\bibitem{eerenstein}
W. Eerenstein, N.D Mathur, J.F. Scott, \emph{Nature}, \textbf{442}, 756--765(2006).

\bibitem{khomskii}
D.~I. Khomskii, \emph{J. Magn. Magn. Mater.}, \textbf{306}, 1--8(2006).

\bibitem{srinivasan}
G. Srinivasan, E.T. Rasmussen, B.J. Levin, R. Hayes, \emph{Phys. Rev. B}, \textbf{65},  134402-134407 (2002)

\bibitem{hur}
N. Hur, S. Park, P.A. Sharma, J.S. Ahn, S. Guha, S-W. Cheong, \emph{Nature}, \textbf{429},  392 (2004)

\bibitem{zheng}
H. Zheng et al., \emph{Science}, \textbf{303}, 661-663(2004).

\bibitem{lee}
S. Lee et al., \emph{Nature}, \textbf{451},  805(2008)

\bibitem{daj}
L.K. Daj, \emph{J. Appl. Phys.}, \textbf{101},  064117--064120(2007)

\bibitem{kenzelmann}
M. Kenzelmann,et al.,\emph{Phys. Rev. Lett.}, \textbf{95}, 087206(2005).



\bibitem{lancaster}
T. Lancaster et al.,\emph{J. Phys.: Condens. Matter}, \textbf{19},  376203(2007)
\bibitem{hill}
N.A. Hill,\emph{J. Phys. Chem. B}, \textbf{104},  6694(2000)


\bibitem{khomskiiP}
D. I. Khomskii, \emph{Physics}, \textbf{2}, 20(2009)

\bibitem{kaul}
E.E. Kaul, N. Haberkorn, J. Guimpel, \emph{Appl. Sur. Science}, \textbf{254}, 160-163 (2007)

\bibitem{belik}
A.A.~Belik, T.~Yokosawa, K.~Kimoto, Y.~Matsui, and E.~Takayama-Muromachi, \emph{Chem. Mater}, \textbf{19},  1679-1689(2007)

\bibitem{montanari}
E.~Montanari, G.~Galestani, L.~Righi, E.~Gilioli, F.~Bolzoni, K.S.~Knight, and Radaelli, \emph{Phys. Rev. B}, \textbf{75}, 220201(2007).
\bibitem{santos}
A.~Moreira dos Santos, I.E.~Cheetham, and  T.~Atou, \emph{Phys. Rev. B}, \textbf{66},  064425(2002)
 
\bibitem{kimura}
T.~Kimura, S.~Kawamoto, I.~Yamada, M.~Azuma, M.~Takano, Y.~Tokura, \emph{Phys. Rev. B}, \textbf{67},  180401(2003)
\bibitem{smol}
G.A.~Smolenskii, and I.E.~Chupis, \emph{Low. Temp. Phys.}, \textbf{137},  415--448(1982)
 
\bibitem{pal}
V. Palchykov, C. von Ferber, R.Folk, R. Yu. Holovatch \emph{Phys. Rev E}, \textbf{80}, 011108 (2009)
\bibitem{chiba}
H.~Chiba, T.~Atou, and Y.~Syono, \emph{J. Solid State Chem.}, \textbf{132},  139-143(1997)

\bibitem{zhong}
C.~Zhong, J.~Fang, and Q.~Jiang, \emph{J. Phys.: Condens. Matter.}, \textbf{16},  9059--9068(2004)

\bibitem{chupits}
I.E.~Chupis, \emph{Usp. Fiz. Nauk}, \textbf{31},  858(2005)

\bibitem{belik2}
A.A.~Belik, and E.~Takayama-Muromachi, \emph{Inorg. Chem.}, \textbf{45},  10224-10229(2006)

\bibitem{strukov}
B.A. Strukov, A.P. Levanyuk, \emph{Ferroelectric Phenomena in Crystals} (Springer-Verlag, Berlin, 1998)

\end{thebibliography}
\end{document}